\documentclass[sn-mathphys,Numbered]{sn-jnl}

\usepackage{amsmath,amssymb,amsfonts}%
\usepackage{xcolor}%
\usepackage{manyfoot}%
\usepackage{booktabs}%
\usepackage[T1]{fontenc}
\usepackage{subcaption}
\usepackage{comment}
\usepackage{lmodern}
\usepackage{lineno}

\definecolor{CentralRegionColor}{HTML}{403990}
\definecolor{FrontalLobeColor}{HTML}{80a6e2}
\definecolor{TemporalLobeColor}{HTML}{fbdd85}
\definecolor{ParietalLobeColor}{HTML}{f46f43}
\definecolor{OccipitalLobeColor}{HTML}{009e73}
\definecolor{LimbicLobeColor}{HTML}{cee000}
\definecolor{SubcorticalGrayNucleiColor}{HTML}{cc79a7}

\begin{document}

\title{Diagnosis and Pathogenic Analysis of Autism Spectrum Disorder Using Fused Brain Connection Graph}


\author[1,5]{\fnm{Lu} \sur{Wei}}

\author[2]{\fnm{Yi} \sur{Huang}}

\author[3]{\fnm{Guosheng} \sur{Yin}}

\author[1]{\fnm{Fode} \sur{Zhang}}

\author[2,4]{\fnm{Manxue} \sur{Zhang}}

\author*[1]{\fnm{Bin} \sur{Liu}}\email{liubin@swufe.edu.cn}

\affil[1]{\orgdiv{Center of Statistical Research, School of Statistics},\orgname{Southwestern University of Finance and Economics}, \orgaddress{\city{Chengdu}, \country{China}}}

\affil[2]{\orgdiv{Mental Health Center}, \orgname{West China Hospital of Sichuan University}, \orgaddress{\city{Chengdu}, \country{China}}}

\affil[3]{\orgdiv{Department of Statistics and Actuarial Science}, \orgname{The University of Hong Kong}, \orgaddress{\city{Hong Kong}, \country{China}}}

\affil[4]{\orgdiv{Mental Health Center}, \orgname{General Hospital of Ningxia Medical University}, \orgaddress{\city{Yinchuan}, \country{China}}}

\affil[5]{\orgdiv{Intelligence and Sensing Lab, School of Information Science and Technology}, \orgname{Osaka University}, \orgaddress{\city{Osaka}, \country{Japan}}}

\abstract{
We propose a model for diagnosing Autism spectrum disorder (ASD) using multimodal magnetic resonance imaging (MRI) data. Our approach integrates brain connectivity data from diffusion tensor imaging (DTI) and functional MRI (fMRI), employing graph neural networks (GNNs) for fused graph classification. To improve diagnostic accuracy, we introduce a loss function that maximizes inter-class and minimizes intra-class margins. We also analyze network node centrality, calculating degree, subgraph, and eigenvector centralities on a bimodal fused brain graph to identify pathological regions linked to ASD. Two non-parametric tests assess the statistical significance of these centralities between ASD patients and healthy controls. Our results reveal consistency between the tests, yet the identified regions differ significantly across centralities, suggesting distinct physiological interpretations. These findings enhance our understanding of ASD's neurobiological basis and offer new directions for clinical diagnosis.
}




\maketitle
\section{Introduction}
Autism spectrum disorder (ASD) is a neurodevelopmental disorder characterized by deficits in social interaction and communication, repetitive behaviors, and restricted interests \cite{Rapin2002NEJM}. Typically manifesting in early childhood, the effects of ASD can become more impactful on a person's life as he/she reaches adulthood. The presentation of symptoms is heterogeneous and varies greatly among individuals. Known etiological factors for ASD include strong genetic components \cite{Sandin2014JAMA}, as well as environmental influences \cite{SEALEY2016288}.

Early diagnosis of ASD is crucial for timely intervention, which can greatly improve outcomes \cite{Johnson2007Pediatrics}. Unfortunately, many children receive a diagnosis at a later age, missing the optimal window for early intervention. Current diagnostic practices primarily rely on observing clinical symptoms, often through specific autism interview \cite{rutter2003autism}. However, these traditional diagnostic methods have two major limitations. First, the high heterogeneity of ASD makes consistent symptom-based clinical diagnosis challenging. Second, the diagnostic process requires highly qualified clinicians, which is time-consuming and labor-intensive. Such highly trained professionals are scarce even in urban hospitals, let alone in rural areas. These issues lead to high rates of misdiagnosis and underdiagnosis of ASD.

Although the pathogenesis of ASD remains inconclusive, existing research indicates a strong association between ASD symptoms and certain neural circuits, such as the corticostriatal circuit, cingulate cortex, and amygdala \cite{2016Persistence}. The human brain is a complex neurophysiological network composed of hundreds of brain regions and thousands of interconnected pathways \cite{2012The}. With medical imaging techniques, it is possible to reconstruct and analyze these brain networks. Magnetic resonance imaging (MRI) is a non-invasive imaging technique that can reveal both the functional and structural aspects of the brain, allowing the reconstruction of structural brain networks. For instance, diffusion tensor imaging (DTI) can be used to calculate the density and efficiency of white matter connections, aiding in the construction of structural brain networks and rich club organizations \cite{van2011rich}. Functional MRI (fMRI), based on the blood oxygen level-dependent (BOLD) contrast principle, can detect changes in neural activity during specific cognitive tasks or stimuli \cite{logothetis2008we}. Using fMRI, researchers can compute group-averaged partial correlation matrices of brain regions and infer significant functional connections with the aid of anatomical templates \cite{Salvador2005}. Specifically, fMRI and DTI can provide two different modalities of brain region networks (90×90) according to AAL-90 template as shown in Table \ref{tab:aal90}, with each of the 90 nodes representing different brain regions. This study aims to fuse these two modalities of brain network data to establish an ASD diagnostic and pathological analysis model. Additionally, we conduct statistical analyses on the 90 brain regions to identify and analyze suspected pathogenic neural circuits from a graph-theoretic perspective.

In particular, we model ASD diagnosis as a graph classification problem, and ASD pathology analysis as the detection of abnormal brain regions within the brain network. Graph neural networks (GNNs) have proven to be highly effective in handling non-Euclidean data and are widely used in various biomedical applications \cite{long2022pre}. To address the issues of ASD diagnosis and abnormal brain region connectivity detection, we adopt a GNN-based diagnostic and pathological analysis framework. Specifically, we use one modality's network as the adjacency matrix and the other modality's network as the node feature matrix, transforming the ASD diagnosis into a binary graph classification problem based on GNNs. To enable the GNNs to capture the differences in brain networks between the ASD group of 67 patients and the control group of 71 individuals from West China Hospital, we propose a maximum margin loss function. This function aims to maximize inter-class distance while minimizing intra-class distance, thus constraining the learning process of graph representation vectors and improving classification performance.

While developing the ASD diagnostic model, we also aim to identify potential pathogenic neural circuits to assist clinicians in further pathological analysis. Specifically, we first use the fused brain network features obtained from the previous classification step to calculate the correlation matrix among the 90 brain regions, resulting in a multimodal fusion brain network. We then calculate the network node centrality measures—degree centrality, eigenvector centrality, and subgraph centrality—for both the ASD and control groups, performing statistical hypothesis tests to identify significant differences. We use two non-parametric tests, Mann-Whitney U (MWU) and maximum mean discrepancy (MMD), to examine the distribution differences of these centrality measures across the 90 brain regions, ultimately identifying candidate brain regions implicated in ASD. Our results lead to the following two conclusions: 1) Different centrality measures yield distinct sets of candidate brain regions; 2) MWU and MMD tests produce consistent results when assessing the distribution differences of the three centrality measures.
To elaborate more on the first conclusion, different network node centrality measures capture the importance of nodes from different perspectives. For instance, subgraph centrality considers a node's importance based on the connectivity of the subgraph it belongs to, particularly closed paths including the node itself, whereas degree centrality is based on the number of edges directly connected to the node. 
We release our code at \url{https://github.com/lobsterlulu/ASD-Diagnosis-Using-Multimodal-MRI.git}.

The contributions of this work can be summarized as follows:
\begin{itemize}
    \item We propose a model that utilizes GNNs to integrate fMRI and DTI imaging data for ASD diagnosis.
    \item We introduce an optimized inter-class and intra-class regularization method to improve diagnostic performance.
    \item Based on the integrated data, we propose a method to test the distribution differences of three different network node centrality measures between the ASD group and the control group, so as to identify abnormal brain regions associated with ASD in order to provide references for subsequent pathological analysis and treatment.
\end{itemize}

\section{Results}
     

\subsection{Bi-modal Data Fusion}
\begin{figure*}[!thp]
    \centering
    \includegraphics[width=13cm]{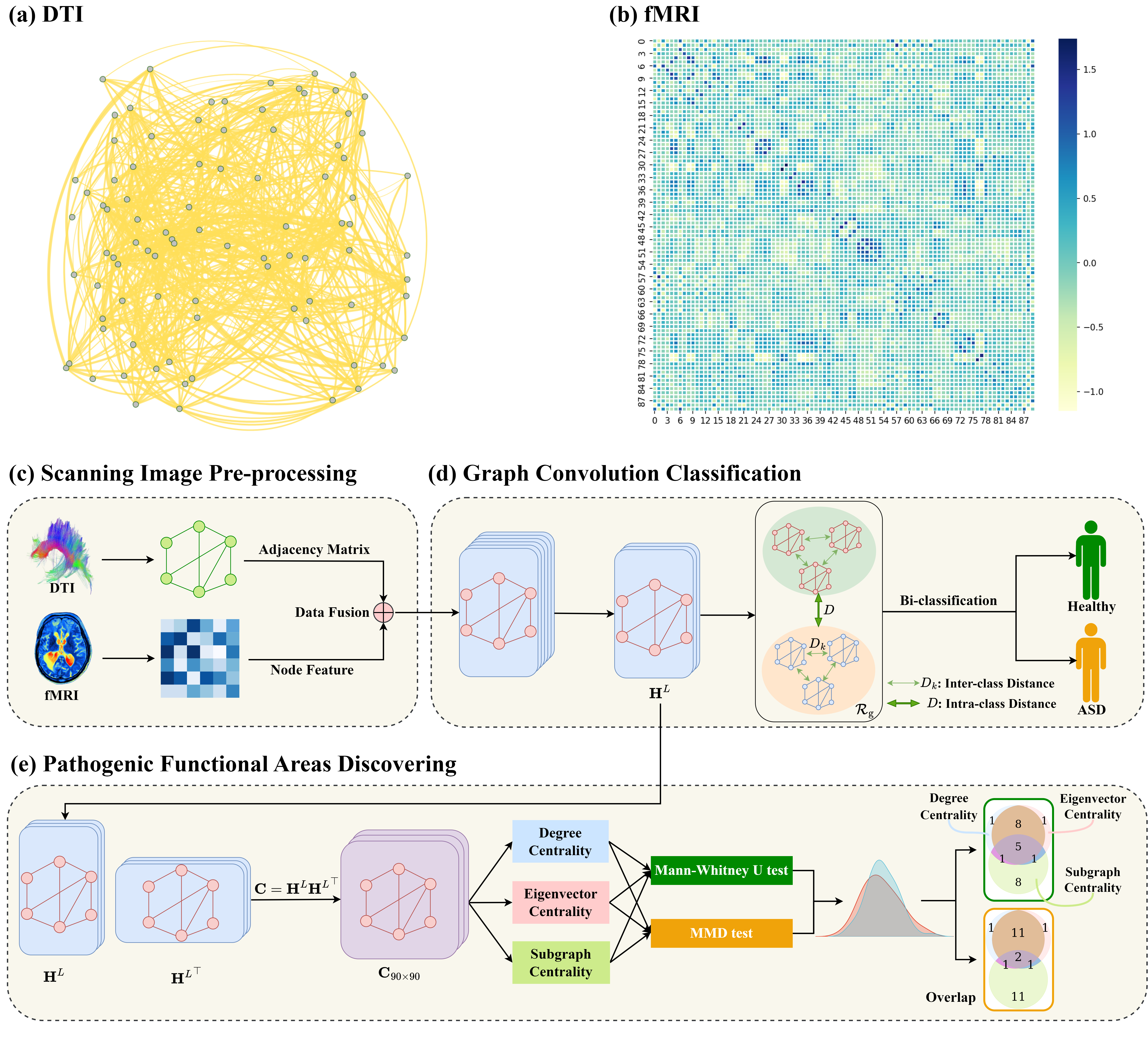}
    \caption{(a)--(b) Visualizations of the brain region adjacency matrices $\mathbf{A}^{\rm DTI}$ and $\mathbf{A}^{\rm fMRI}$ obtained from DTI (a) and fMRI (b) modalities, respectively. In panel (a), the nodes represent the 90 brain regions, while in panel (b), the columns/rows correspond to the same set of brain regions. The workflow of the proposed model with three modules panel (c)--(e).}
    \label{fig:modelFramework}
\end{figure*}
DTI enables noninvasive {\em in vivo} visualization of white matter fibers, facilitating the study of brain development and functional cognition. In contrast, fMRI has high spatial resolution and leverages changes in blood oxygen concentration to obtain functional signals that reflect local brain activity, enabling the detection of abnormalities in language and cognitive abilities. We construct two adjacency matrices $\mathbf{A}^{\rm DTI}, \mathbf{A}^{\rm fMRI}\in \mathbb{R}^{90\times 90}$ to represnt the DTI and fMRI brain region connection graphs as detailed in Sections \ref{sec:dtiPreprocessing} and \ref{sec:fmriPreprocessing}, respectively. Figures \ref{fig:modelFramework} (a) and \ref{fig:modelFramework} (b) visualize the DTI and fMRI of one patient. 

As the functions and structures of the human brain are intricately linked, multiple functional regions must collaborate to perform various tasks, with specific functional regions requiring specific anatomical connections. This study uses the fusion of brain functional structure map data obtained from DTI and fMRI to improve ASD identification.
We fuse the bimodal graphical data $\mathbf{A}^{\rm DTI}$ and $\mathbf{A}^{\rm fMRI}$ using GNNs. The principle behind GNNs is to perform graph convolution operations on node features using the adjacency matrix of the graph. Specifically, a GNN operation, denoted as ${\rm GNN}(\mathbf{A},\mathbf{X})$, takes the adjacency matrix $\mathbf{A}$ and node features $\mathbf{X}$ of a graph as the input.
To fuse the DTI and fMRI data within the framework of GNNs, we can choose one of the matrices, $\mathbf{A}^{\rm DTI}$ and $\mathbf{A}^{\rm fMRI}$, as the adjacency matrix $\mathbf{A}$ and the other matrix as the node features $\mathbf{X}$. Figure \ref{fig:modelFramework} (c) demonstrates the solution for ${\rm GNN}(\mathbf{A}=\mathbf{A}^{\rm DTI},\mathbf{X}=\mathbf{A}^{\rm fMRI})$, where we use DTI data as the adjacency matrix and fMRI data as the node feature. Alternatively, we can exchange $\mathbf{A}^{\rm DTI}$ and $\mathbf{A}^{\rm fMRI}$ in ${\rm GNN}(\cdot,\cdot)$ to have ${\rm GNN}(\mathbf{A}=\mathbf{A}^{\rm fMRI},\mathbf{X}=\mathbf{A}^{\rm DTI})$. 
Subsequently, we engage in graph classification for ASD diagnosis utilizing the representation ${\rm GNN}(\mathbf{A},\mathbf{X})$. In order to enhance performance, we introduce a regularization term designed to regulate both inter-class and intra-class distances, as illustrated in Figure \ref{fig:modelFramework} (d). For further elaboration, please consult the Method section.

To evaluate the effectiveness of our proposed method, we employ four common graph convolutions, namely graph convolutional network (GCN) \citep{kipf2017semi}, graph attention network (GAT) \citep{veličković2018graph}, ChebyNet \citep{Defferrard2016ChebyNet}, GraphSAGE \citep{Hamilton2017Graphsage}, as the aggregation operation in Eq. (\ref{eq:graphconvol}). We adopt the Adam optimizer with a weight decay of $1\times 10^{-3}$ to train our models, while the batch size and learning rate are set to be 68 and $8\times 10^{-3}$, respectively. To balance the contributions of negative and positive samples for the regularization term in Eq.~(\ref{eq:graphRegularization}), we propose to draw a similar number of positive and negative samples to form a batch. Moreover, we fix the trade-off factor $\alpha$ in Eq.~(\ref{eq:overallLoss}) for all models. Finally, we conduct 10-fold cross-validation to evaluate the models.

\begin{figure}[!htp]
    \centering    \includegraphics[width=13cm]{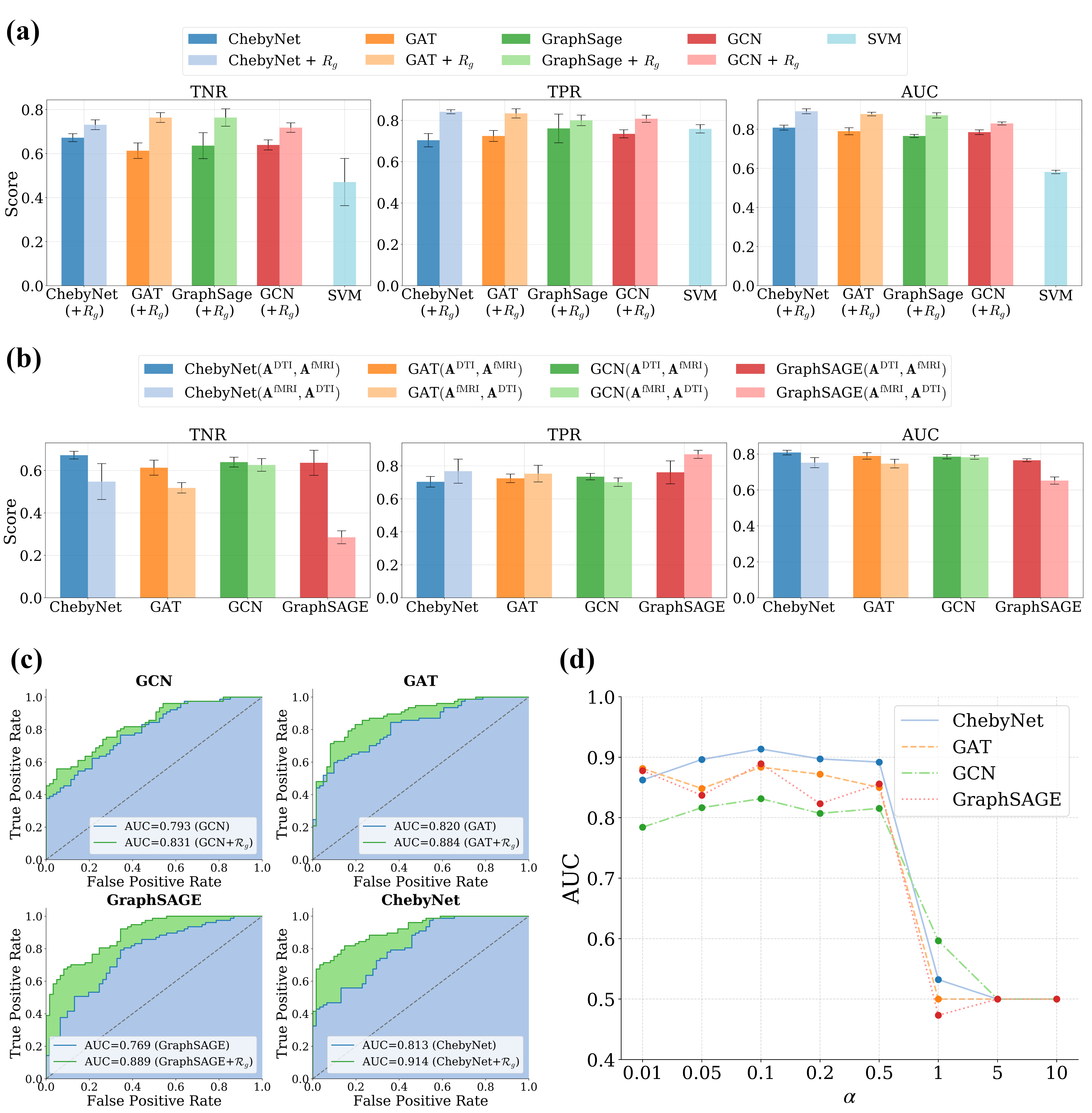}
    \caption{(a) Visualizations of ASD classification on TNR (left panel), TPR (middle panel), and AUC (right panel) (with standard deviation) of baseline methods and the proposed models, averaged over five cross-validation folds. ``(+ $\mathcal{R}_{g}$)'' denotes the proposed regularization-based GNN model. (b) Comparisons on two fusion manners, ${\rm GNN}(\mathbf{A}=\mathbf{A}^{\rm DTI},\mathbf{X}=\mathbf{A}^{\rm fMRI})$ and  ${\rm GNN}(\mathbf{A}=\mathbf{A}^{\rm fMRI},\mathbf{X}=\mathbf{A}^{\rm DTI})$, namely, exchange the roles of $\mathbf{A}=\mathbf{A}^{\rm DTI},\mathbf{X}=\mathbf{A}^{\rm fMRI}$ in GNNs, where ${\rm GNN}$ can be selected from \{GCN, GAT, GraphSAGE, 
ChebyNet\}. (c) ROC curves of the four baselines (blue) and the corresponding proposed methods (green). (d) Searching for the penalty tuning parameter $\alpha$ in Eq.~(\ref{eq:overallLoss}) of the four proposed models.}
    \label{fig:poltall_figs}
\end{figure}

We have five baseline models: GCN, GAT, GraphSAGE, and ChebyNet models without the $\mathcal{R}_{g}$ term as shown in Eq.~(\ref{eq:overallLoss}). In addition, we choose Support Vector Machine (SVM) as another baseline, which accepts concatenated bimodal matrices $\mathbf{A}^{\rm DTI}_i \oplus \mathbf{A}^{\rm fMRI}_i$ as input features (where $\oplus$ denotes the concatenate operation).

We report the true negative rate (TNR), true positive rate (TPR), and area under the curve (AUC) to evaluate the classification performance. The overall results of the baselines and the proposed model are presented in Figure \ref{fig:poltall_figs} (a).
We observe that our proposed optimal class distance GNN-based model exhibits significant performance improvement in all three evaluation metrics compared with the four base models. This indicates that the regularization term $\mathcal{R}_{g}$ provides substantial advantages for prediction. 
In addition, we observe that SVM performs the worst in this problem, while ChebyNet$+\mathcal{R}_{g}$ achieves the best AUC. The average AUC of the model ChebyNet$+\mathcal{R}_{g}$ is 0.8917, which is 10\% higher than the baseline model. Furthermore, GAT$+\mathcal{R}_{g}$ achieves the best overall performance in terms of TPR, TNR, and AUC.

Figure \ref{fig:poltall_figs} (c) visualizes the receiver operating characteristic (ROC) curves of the optimal prediction results of the eight GNNs-based methods. The highest single AUC of 0.914 is achieved by ChebyNet$+\mathcal{R}_{g}$. The biggest improvement over the original model is observed in the GraphSAGE model, with a nearly 14\% improvement. The experimental results demonstrate that the introduction of the regularization term $\mathcal{R}_{g}$ with Wasserstein distance improves the classification performance of the model.


\subsubsection{Hyperparameter Search}
Figure \ref{fig:poltall_figs} (d) illustrates the process of searching for the best penalty regularization parameter $\alpha$ in Eq.~(\ref{eq:overallLoss}). We select $\alpha$ based on the final AUC, and the candidate set of $\alpha$ is $\{0.01, 0.05, 0.1, 0.2, 0.5, 1.0, 5, 10\}$. We observed that all our four models achieved the highest AUC at $\alpha=0.1$, which indicates that our models are robust in the choice of hyperparameter $\alpha$. This also implies that the regularization term $\mathcal{R}_{g}$ has a consistent impact on the performance across different $\alpha$ values.

\subsubsection{Alternative Fusion Mode}
\label{sec:alternativeFusionOption}

In the bimodal data fusion process, we designate one of the matrices, $\mathbf{A}^{\rm DTI}$ or $\mathbf{A}^{\rm fMRI}$, as the adjacency matrix $\mathbf{A}$, while the other serves as the node features $\mathbf{X}$, which results two solutions, ${\rm GNN}(\mathbf{A}=\mathbf{A}^{\rm DTI},\mathbf{X}=\mathbf{A}^{\rm fMRI})$ and ${\rm GNN}(\mathbf{A}=\mathbf{A}^{\rm fMRI},\mathbf{X}=\mathbf{A}^{\rm DTI})$. In practice, we can try different ${\rm GNN}$ models, such as GCN, GAT, GraphSAGE, and
ChebyNet.
Figure \ref{fig:poltall_figs} (b) demonstrates the TNR, TPR, and AUC of two fusion methods as shown in the left, middle, and right panels respectively. 
There is no significant difference in stability between the two fusion methods. 
However, in terms of AUC, TPR, and TNR, we observe that the fused data of ${\rm GNN}(\mathbf{A}=\mathbf{A}^{\rm DTI},\mathbf{X}=\mathbf{A}^{\rm fMRI})$ outperforms ${\rm GNN}(\mathbf{A}=\mathbf{A}^{\rm fMRI},\mathbf{X}=\mathbf{A}^{\rm DTI})$ for each model in terms of prediction. Among them, the AUC of ${\rm GCN}(\mathbf{A}=\mathbf{A}^{\rm fMRI},\mathbf{X}=\mathbf{A}^{\rm DTI})$ is the highest at 0.7850, while the AUC of ${\rm GraphSAGE}(\mathbf{A}=\mathbf{A}^{\rm fMRI},\mathbf{X}=\mathbf{A}^{\rm DTI})$ is the lowest at 0.6524. This result has the largest difference from ${\rm GNN}(\mathbf{A}=\mathbf{A}^{\rm DTI},\mathbf{X}=\mathbf{A}^{\rm fMRI})$ among all models.


\subsection{Pathogenic Analysis}
\label{sec:pathAna}

Apart from ASD diagnosis, clinicians express interest in pathological analysis with the objective of pinpointing the brain regions or connections linked to ASD. We carry out this analysis utilizing the integrated patient representation $\mathbf{H}^L$. Consequently, we compute the fused correlation matrix $\mathbf{C}= \mathbf{H}^L {\mathbf{H}^L}^{\top}\in \mathbb{R}^{90\times 90}$ for all samples, as depicted in Figure \ref{fig:modelFramework} (e).

\begin{figure}[!thp]
    \centering
    \includegraphics[width=13cm]{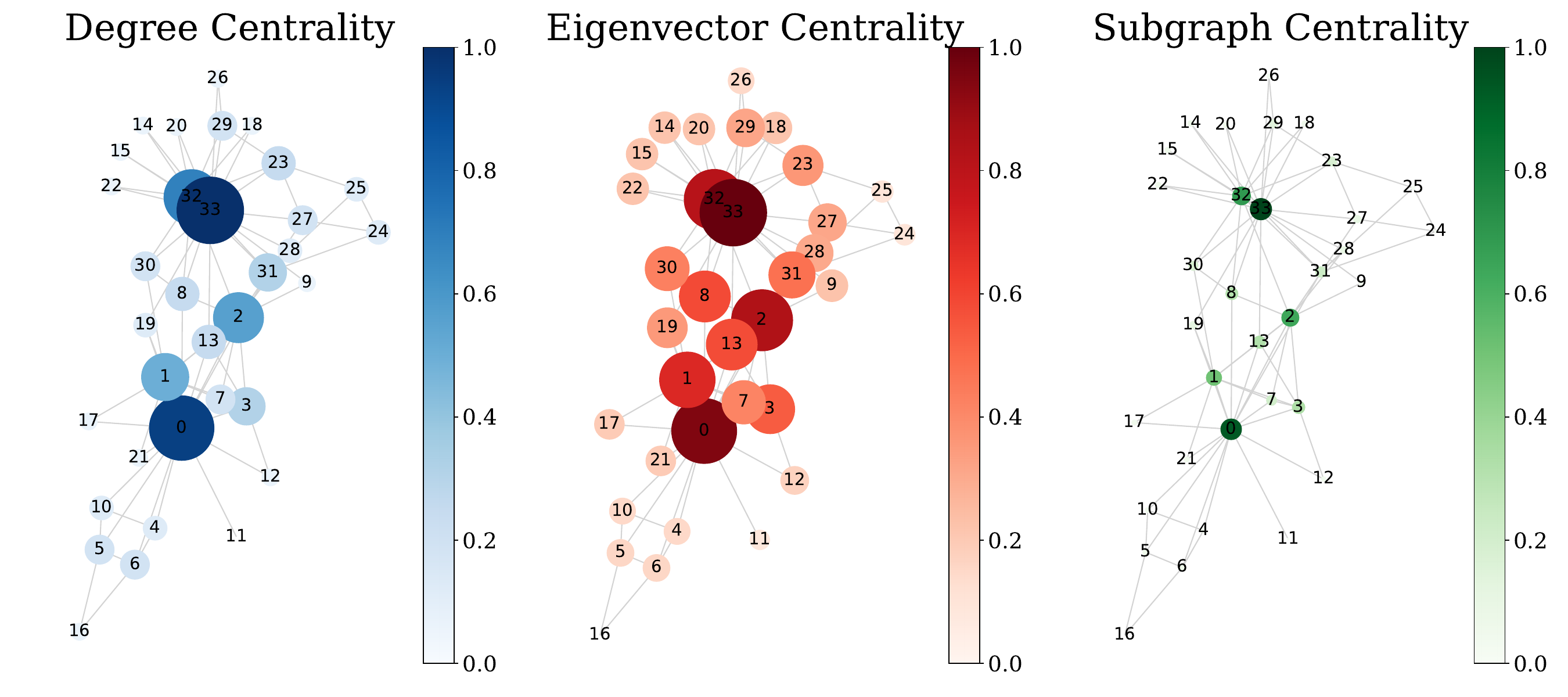}
    \caption{The visualization depicts three distinct centralities of the Karate Club graph: degree centrality (left panel), eigenvector centrality (middle panel), and subgraph centrality (right panel). The color intensity and node size correspond to normalized centrality values, with larger and darker nodes indicating higher centrality levels.}
    \label{fig:centralities}
\end{figure}

The matrix $\mathbf{C}$ serves as a fused graph that comprehensively represents the brain. We evaluate its node centrality across positive and negative groups to identify anomalous nodes (brain regions). 
One approach involves detecting anomalous nodes by leveraging the node centrality of $\mathbf{C}$ exemplified as follows. 
We propose to perform an overall pathogenic analysis with three types of network node centrality:
degree centrality,
eigenvector centrality,
and subgraph centrality,
as shown in Figure~\ref{fig:centralities}. Section \ref{sec:graphNodeCentrality} summarizes the details about the three types of network node centrality.

Taking the eigenvector centrality as an example, we first calculate all 90 eigenvector centrality values of $\mathbf{C}$, denoted as ${\lambda_1, \ldots, \lambda_{90}}$. It should be noted that all the eigenvector centrality values of $\mathbf{C}$ are real since it is a semi-definite matrix. We can divide all 138 samples into two groups: the ASD group ($+$) and the control group ($-$). Therefore, the $i$-th eigenvector centrality value $\lambda_i$ can also be divided into the ASD and control groups,
\begin{equation*}
\begin{aligned}
    S_i^{+}&:=\left \{\lambda_{i,j}^{+}\right \}_{j=1}^{N_{\rm ASD}} \quad \text{vs} \quad
    S_i^{-}&:=\left \{\lambda_{i,j}^{-}\right \}_{j=1}^{N_{\rm Ctr}},
\end{aligned}
\end{equation*}
where $N_{\rm ASD}=67$ and $N_{\rm Ctr}=71$ are the sizes of ASD group and control group respectively.

We then perform a hypothesis test on the distribution of the eigenvector centrality values of $\mathbf{C}$ to find the contaminated nodes for ASD. Specifically, we propose the null hypothesis as positive and negative samples follow the same distribution,
\begin{equation}\label{eq:hypothesisTest}
    H_0: P(S_i^{+}) = P(S_i^{-})
\end{equation}
where $P(S_i^{+})$ and $P(S_i^{-})$ denote the distributions of the $i$-th eigenvector centrality value in the ASD group and control group, respectively.

To enhance the comprehensiveness of the pathological analysis, not only do we incorporate three distinct network centrality measures, but we also utilize two different non-parametric testing methodologies, namely the MWU test and the MMD test, to conduct the hypothesis testing outlined in Eq.~(\ref{eq:hypothesisTest}). The MWU and MMD tests are detailed in Section \ref{sec:hypothesisTestDefinitions}.

\subsubsection{Results of Pathogenic Analysis}
\begin{figure*}[!hpt]
    \centering
    \includegraphics[width=13cm]{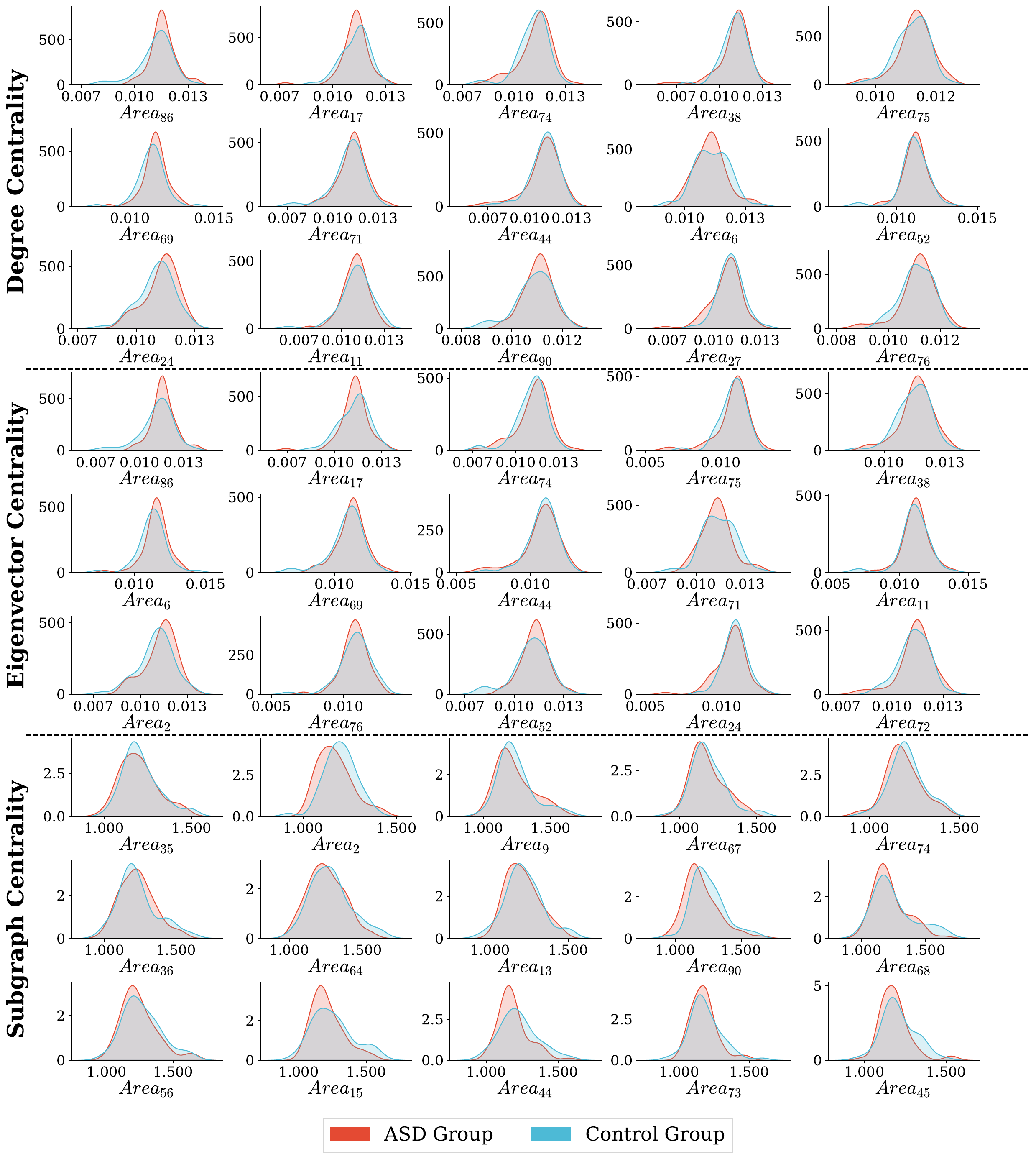}
    \caption{The visualization portrays the distributions of the three types of network node centrality within the ASD group (comprising 67 patients denoted by red curves) and the control group (consisting of 71 samples denoted by blue curves). The top 15 distributions of brain regions are displayed based on the ranking of network node centrality. The panels, from top to bottom, correspond to degree centrality, eigenvector centrality, and subgraph centrality, respectively.
    }
    \label{fig:hyposistestDistri}
\end{figure*}

For pathological analysis, our focus involves filtering ASD-associated biomarkers through the examination of the distributions pertaining to three network node centralities: degree centrality, eigenvector centrality, and subgraph centrality of $\mathbf{C}$ across both ASD and control groups. Subsequently, we conduct experiments employing both the MWU and MMD tests.

Taking the MMD test as an illustration, we compute the MMD loss between the ASD group and the control group across all 90 brain regions for the three types of node centrality. Subsequently, we sort the 90 brain regions based on their MMD loss values.
Figure \ref{fig:hyposistestDistri} illustrates the distributions of the ASD group (depicted by red curves) and the control group (represented by blue curves) for the top 15 brain regions ranked by the MMD loss, categorized by degree centrality (top panel), eigenvector centrality (middle panel), and subgraph centrality (bottom panel).
For instance, in the middle panel, the first sub-figure displays the distributions of the 86-th brain region's ($Area_{86}$) eigenvector centrality for the ASD group and the control group respectively. Notably, the disparity concerning $Area_{86}$ emerges as the most pronounced among the 90 brain regions, suggesting $Area_{86}$ to be the most suspicious brain region for ASD based on the eigenvector centrality perspective.

From the insights gleaned in Figure \ref{fig:hyposistestDistri}, it is apparent that the top 15 brain regions identified by both degree centrality and eigenvector centrality exhibit substantial overlap. However, these two sets display minimal overlap with the brain regions identified by subgraph centrality. This observation aligns with the rationale depicted in Figure \ref{fig:centralities}.


\begin{figure*}[!htp]
    \centering
    \includegraphics[width=13cm]{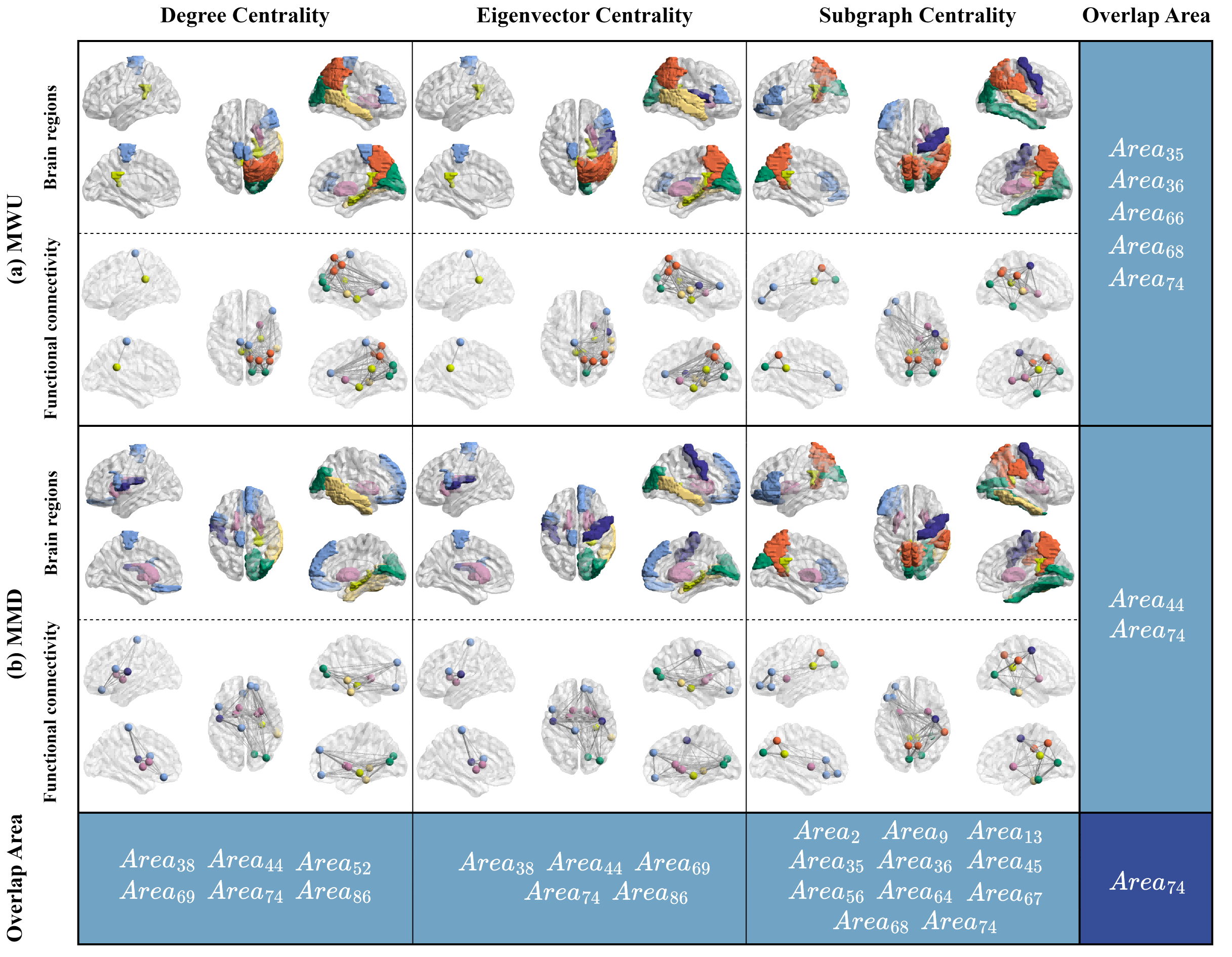}
    \caption{Visualization of ASD-related brain regions and functional connectivity related to ASD, analyzed through two statistical methods: (a) MWU test (rows 1 and 2) and (b) MMD test (rows 3 and 4). Three centrality measures are displayed for each testing method: degree centrality (column 1), eigenvector centrality (column 2), and subgraph centrality (column 3). Rows 1 and 2 show brain regions and functional connectivity for both left and right hemispheres of the brain. Each sub-figure, for example, the figure of row 1 and column 1, shows significant regions of the left and right halves of the brain from lateral and dorsal views, and the medial view visualizes the top 15 ASD-related regions. The brain is divided into 7 regions, each represented by a different color in the subgraphs: 
    \textcolor{CentralRegionColor}{Central Region}, \textcolor{FrontalLobeColor}{Frontal Lobe}, \textcolor{TemporalLobeColor}{Temporal Lobe}, \textcolor{ParietalLobeColor}{Parietal Lobe}, \textcolor{OccipitalLobeColor}{Occipital Lobe}, \textcolor{LimbicLobeColor}{Limbic Lobe}, and \textcolor{SubcorticalGrayNucleiColor}{Sub Cortical Gray Nuclei}.
    The overlapping areas between different centrality measures are indicated in the rightmost column while different test methods are indicated at the bottom row in light blue. The dark blue area highlights the final intersection of regions that consistently overlap across all centrality measures and statistical tests.}
    \label{fig:visualization_brainRegion_mwu}
\end{figure*}


Figure~\ref{fig:visualization_brainRegion_mwu} illustrates the prominent brain regions (rows 1, 3) and functional connectivity (rows 2, 4) associated with ASD as determined by two statistical tests: the MWU test in panel (a) and the MMD test in panel (b). Columns 1, 2, and 3 of Figure~\ref{fig:visualization_brainRegion_mwu} correspond to the Degree, Eigenvector, and subgraph centrality. Each subfigure within the columns showcases the brain from three distinct perspectives: the left and right images reveal significant brain regions from lateral and dorsal views, emphasizing the left and right hemispheres, respectively. The central image displays the brain from a medial view, highlighting the top 15 identified significant regions. 
The visualization of brain regions and functional connectivity serves as different representations of the same result, focusing respectively on regions and connections. To facilitate a more effective comparison of overlapping areas between different methods, the brain is segmented into seven regions distinguished by unique colors: central region, frontal lobe, temporal lobe, parietal lobe, occipital lobe, limbic lobe, and subcortical gray nuclei, following the segmentation method outlined in \cite{TZOURIOMAZOYER2002273}.

The MWU test returns both effect sizes and p-values, which are used to quantify the importance of the 90 nodes (brain regions). Based on the effect sizes, we select the top 15 important nodes as candidate ASD imaging biomarkers. Panel (a) of Figure~\ref{fig:visualization_brainRegion_mwu} illustrates the results of the MWU test, where we observe a large overlap between the 15 nodes calculated using degree centrality and eigenvector centrality. There are 13 common nodes include \{$Area_{69}$: PCL L, $Area_{86}$: MTG R, $Area_{36}$: PCG R, $Area_{38}$: HIP R, $Area_{74}$: PUT R, $Area_{44}$: CAL R, $Area_{68}$: PCUN R, $Area_{35}$:PCG L, $Area_{60}$: SPG R, $Area_{46}$: CUN R, $Area_{14}$: IFGtriang R, $Area_{62}$: IPL R, $Area_{66}$: ANG R\}, which are concentrated in the right hemisphere. However, after introducing the overlap with subgraph centrality, the common intersection of all three centrality methods becomes 5 areas shown in panel (a) in light blue on the right, specifically \{$Area_{35}$:PCG L, $Area_{36}$: PCG R, $Area_{66}$: ANG R, $Area_{68}$: PCUN R, $Area_{74}$: PUT R\}. Similarly, for the MMD test as shown in Panel (b) of Figure~\ref{fig:visualization_brainRegion_mwu}, the intersection of the top 15 significant ASD imaging biomarkers across the three centralities are \{$Area_{44}$: CAL R, $Area_{74}$: PUT R\}. We see that the brain regions with more overlap of important nodes in the three imaging views are in the prefrontal cortex, including the Lenticular nucleus ($Area_{74}$), calcarine fissure and surrounding cortex ($Area_{44}$), and paracentral lobule ($Area_{69}$), suggesting a focus on motor control, higher cognitive functions, and visual processing areas. 

For degree centrality and eigenvector centrality, there exists a notable overlap among the top 15 regions identified through two distinct testing methods, depicted in the blue blocks of the last row in Figure~\ref{fig:visualization_brainRegion_mwu}. In contrast, subgraph centrality reveals a different intersection but showcases a relatively higher number of overlapping regions, totaling eleven, which include: \{$Area_{2}$: PreCG R, $Area_{9}$: ORBmid L, $Area_{13}$: IFGtriang L, $Area_{35}$: PCG L, $Area_{36}$: PCG R, $Area_{45}$: CUN L, $Area_{56}$: FFG R, $Area_{64}$: SMG R, $Area_{67}$: PCUN L, $Area_{68}$: PCUN R, $Area_{74}$: PUT R \}. Notably, the region common across all methods is highlighted in dark blue, corresponding to $Area_{74}$: PUT R.

Based on the findings depicted in Figure~\ref{fig:visualization_brainRegion_mwu}, the first observation is that the two hypothesis testing methods yield consistent results. Furthermore, there is another interesting observation: the three types of centralities offer distinct insights into ASD. While degree and eigenvector centralities yield akin conclusions, they diverge from the subgraph centrality findings. This disparity arises from the unique nature of each centrality measure. 
From a clinical perspective, due to the considerable heterogeneity in the pathogenesis of ASD itself \cite{mottron2020autism,lenroot2013heterogeneity}, the pathological analysis results may manifest in various network node centralities. Despite the inconsistencies, the proposed method identifies the pathological brain region $Area_{74}$: PUT R (lenticular nucleus, putamen of the right hemisphere), which is jointly selected by three types of network centrality measures. 
Numerous existing studies have demonstrated the association of the right putamen (PUT R) with ASD \cite{schuetze2016morphological,van2018cortical,del2022neuroanatomical}. In the study by Del et al. (2022) \cite{del2022neuroanatomical}, the authors specifically identify the right putamen's involvement in social cognitive functions, a domain known to be particularly affected in individuals with ASD, closely mirroring our findings. Other partially intersecting regions such as the left Precuneus ($Area_{67}$) and right Precuneus ($Area_{68}$) have also been identified to be associated with ASD \cite{kitamura2021association,silk2006visuospatial}.

\section{Discussion}
Early screening for autism is crucial for effective treatment. This paper introduces an autism screening system based on the fusion of fMRI and DTI imaging. Not only does the system accurately diagnose ASD, but it also identifies suspicious brain regions potentially linked to the disorder through network node centrality analysis, aiding doctors in pathology analysis. Our screening model was trained and tested on a dataset of brain networks from 138 samples, each containing two modalities of data derived from preprocessed fMRI and DTI images. The data fusion strategy involves using the brain network from one modality as the adjacency matrix and that from the other modality as the features of the nodes (brain regions). By exchanging the roles of the modalities, two different models are obtained. Validation results on both models meet clinical screening requirements.

To enhance the classification performance, we add regularization terms to the GNNs for constraining intra-class and inter-class distances. The highest achieved AUC is 91.4\%. Using ten-fold cross-validation, we achieve an average AUC of 89.17\%, with a corresponding specificity (TNR) of 73.11\% and a sensitivity (TPR) of 84.16\%. 
Notably, the classification performance is improved by approximately 7\% overall, and up to 14\% in some cases, after adding the classification constraints.
Our study is still ongoing, and we are currently gathering more patient data. In the future, we plan to increase both training and testing samples to obtain a more stable classification model.

To further enhance the screening performance of our model, we propose a method to identify suspicious brain regions based on the fused network graph of 90 brain regions using three different network node centralities: degree centrality, eigenvector centrality, and subgraph centrality. We then perform statistical hypothesis testing on these centralities using two different non-parametric methods, retaining the top 15 brain regions with significant scores as candidate pathogenic regions. The results demonstrate that the intersection of suspicious brain regions identified through two different hypothesis tests on degree centrality is \{$Area_{38}$: HIP R, $Area_{44}$: CAL R, $Area_{52}$: MOG R, $Area_{69}$: PCL L, $Area_{74}$: PUT R, $Area_{86}$: MTG R\}. The intersection for eigenvector centrality is \{$Area_{38}$: HIP R, $Area_{44}$: CAL R, $Area_{69}$: PCL L, $Area_{74}$: PUT R, $Area_{86}$: MTG R\}, and for subgraph centrality, it is \{$Area_{2}$: PreCG R, $Area_{9}$: ORBmid L, $Area_{13}$: IFGtriang L, $Area_{35}$: PCG L, $Area_{36}$: PCG R, $Area_{45}$: CUN L, $Area_{56}$: FFG R, $Area_{64}$: SMG R, $Area_{67}$: PCUN L, $Area_{68}$: PCUN R, $Area_{74}$: PUT R\}. The intersection of the top 15 significant brain regions for the MWU test across all three centralities is \{$Area_{35}$: PCG L, $Area_{36}$: PCG R, $Area_{66}$: ANG R, $Area_{68}$: PCUN R, $Area_{74}$: PUT R\}, and for the MMD test, it was \{$Area_{44}$: CAL R, $Area_{74}$: PUT R\}. It is important to note that different network node centralities characterize nodes from different perspectives; hence, the results from these three centralities do not completely overlap, providing varied evidence for ASD pathology analysis from a network science perspective.
This disparity can be explained by the fact of the high heterogeneity of the pathogenesis of ASD \cite{mottron2020autism,jiang2022signalling,zhuang2024autism}.

Despite the inconsistencies, the proposed method identifies the pathological brain region $Area_{74}$: lenticular nucleus, putamen of the right hemisphere, which is concurrently selected by three types of network centralities. The putamen has been linked to ASD in previous literature~\cite{schuetze2016morphological,van2018cortical}. In a study by Del et al. (2022)~\cite{del2022neuroanatomical}, the authors further revealed a volumetric reduction in the right putamen associated with social cognitive functions, particularly impaired in ASD, aligning precisely with our findings.

Additionally, other partially overlapping regions, such as the CAL R region ($Area_{44}$) and MOG R region ($Area_{52}$), which are located within the occipital lobe, have been identified as having volumes that correlate with ASD \cite{piven1996regional, carper2002cerebral}. Furthermore, both the left Precuneus ($Area_{67}$) and right Precuneus ($Area_{68}$) have also been found to be associated with ASD \cite{kitamura2021association, silk2006visuospatial}.


In clinical practice, ASD screening primarily relies on experienced doctors using questionnaires. Currently, in China, only top-tier hospitals are equipped with qualified psychiatrists. This issue is particularly severe in rural areas, where ASD in adolescents is often overlooked as a problem of incomplete brain development. MRI-based screening technology can assist rural doctors in quickly screening for ASD, facilitating the transfer of patients to top hospitals for appropriate treatment.

However, the current research has some limitations. First, the medical interpretation of different network node centralities in the brain is unclear, and the contribution of conclusions based on these centralities to ASD pathology analysis needs further clinical validation. Second, our screening model has not been tested on sufficiently large external validation sets. 
Therefore, future work involves collecting more cases that meet these criteria for broader screening applications.

\section{Methods}
\subsection{Materials}
The dataset used in this study consists of 138 participants from west China hospital, with 61 ASD patients and 77 control subjects. 
All participants completed the clinical assessment and MRI scans. The ASD assessment process aims to determine whether an individual meets the diagnostic criteria for ASD as listed in DSM-5. The assessment includes two standardized diagnostic tools: the Autism Diagnostic Observation Schedule (ADOS)~\cite{lord2000autism} and the Autism Diagnostic Interview-Revised (ADI-R)~\cite{lord1994autism}. ADOS is based on a 30--45 minute direct assessment through play, while ADI-R involves a 1.5--2.5 hour interview with the parents. These tools provide a comprehensive evaluation of an individual’s social interaction, communication and repetitive behaviors.

Each subject has bimodal DTI and fMRI data, providing a comprehensive view of the brain activity and structure.
Multimodal MRI data, including fMRI and DTI, are analyzed to provide both qualitative and quantitative information about the brain's structure and function. This information can be used to identify abnormal brain development and neurological features associated with ASD at an early stage of the disease. 

\subsubsection{DTI Preprocessing}
\label{sec:dtiPreprocessing}

While MRI employs magnetic fields and radiofrequency pulses to generate images of body tissues, DTI supplements this process with gradient pulses sensitive to molecular diffusion. This addition enables the measurement and analysis of water molecule diffusion properties within tissues. Given that the human body consists of roughly 70\% water, the random movement of water molecules within brain tissues yields diffusion coefficients that DTI can quantify. Notably, DTI distinguishes between normal individuals and patients based on disparities in diffusion coefficients. The diffusion phenomenon is captured by a three-dimensional tensor $\mathbf{D}$. Due to the symmetric nature of the diffusion matrix $\mathbf{D}$, it suffices for analysis by acquiring a diffusion-weighted image in a minimum of six directions ($\mathbf{D}_{xx}$, $\mathbf{D}_{xy}$, $\mathbf{D}_{xz}$, $\mathbf{D}_{yy}$, $\mathbf{D}_{yz}$, $\mathbf{D}_{zz}$) 
alongside an image devoid of diffusion sensitivity (i.e., a diffusion sensitivity factor) \cite{van2011rich}.
\begin{equation}
\mathbf{D}=\begin{Bmatrix}
\mathbf{D}_{xx} & \mathbf{D}_{xy} & \mathbf{D}_{xz}\\
\mathbf{D}_{yx} & \mathbf{D}_{yy} & \mathbf{D}_{yz}\\
\mathbf{D}_{zx} & \mathbf{D}_{zy} & \mathbf{D}_{zz}
\end{Bmatrix}
\end{equation}

Upon diagonalizing the matrix $\mathbf{D}$, the eigenvalues corresponding to each of the three spatial coordinate axes can be readily derived: $\lambda_1$, $\lambda_2$ and $\lambda_3$, with $\bar{\lambda} = \lambda_1 + \lambda_2 + \lambda_3$. These eigenvalues represent the diffusion rates along three axes, when eigenvalues exhibit proximity, water molecule dispersion manifests isotropic behavior; conversely, distinct values suggest anisotropic characteristics \cite{le2001diffusion}. Anisotropy serves as a scalar depiction of the extent of asymmetric water molecule diffusion within voxels, with fractional anisotropy (FA) being a prevalent scalar metric ranging from 0 to 1. Computed from the eigenvalues of the diagonalized diffusion tensor, FA serves as a measure of diffusion anisotropy, offering insights into the level of microstructural order or organization within white matter regions,
\begin{equation}
    {\rm FA}=\sqrt{\frac{
    3 \left[ {\left( \lambda_1-\bar{\lambda} \right)}^2+
    {\left( \lambda_2-\bar{\lambda} \right)}^2+
    {\left( \lambda_3-\bar{\lambda} \right)}^2\right]}
    {2 \left( {\lambda_1}^2+{\lambda_2}^2+{\lambda_3}^2 \right)}}.
    \end{equation}

The preprocessing of DTI data comprises two primary stages: structural network reconstruction and region segmentation. The initial step in brain network reconstruction involves rectifying potential susceptibility distortions in DTI images using distortion maps. Subsequently, the diffusion weighted images encompassing 33 gradient directions for each subject are aligned with their corresponding images. The estimation of the diffusion tensor model is then conducted via the DTIFIT 
command\footnote{\url{https://fsl.fmrib.ox.ac.uk/fsl/docs/\#/diffusion/dtifit}} in FMRIB's Software Library \cite{jenkinson2012fsl} to derive tensors. Ultimately, the FACT (Fiber Assignment by Continuous Tracking) algorithm \cite{mori2002fiber} is applied to enable deterministic fiber tracking based on eigenvalues and individual anisotropy scores, culminating in the reconstruction of the white matter tracts within the brain network.

The regional segmentation step utilizes the AAL 90 template\footnote{A digital brain structure atlas created by Tzourio-Mazoyer \cite{TZOURIOMAZOYER2002273} for identifying active brain regions, endorsed by the Montreal neurological institute.} as shown in Table \ref{tab:aal90}. This process involves partitioning a subject's brain into 90 regions based on the spatial brain atlas. 
Following regional segmentation, image registration is conducted between the individual's $T1$ images and DTI images. Subsequently, the aligned individual DTIs are projected onto the $T1w$ phase of the Montreal neurological institute standard space. The integration of DTI and anatomical template data is then employed to determine the total number of fiber-connected region pairs $i$ and $j$, resulting in the computation of the connection weight $[\mathbf{A}^{\rm DTI}]_{ij}$. This operation yields an adjacency matrix $\mathbf{A}^{\rm DTI}\in \mathbb{R}^{90\times 90}$, characterizing the DTI data graph.

\begin{table}[!htp]
\centering
\scalebox{0.96}{
\begin{tabular}{@{}cll@{}}
\toprule
Index & Regions abbreviations & Brain regions (full name) \\ \midrule
1/2 & PreCG L/R & Precentral gyrus          \\
3/4 & SFGdor L/R & Superior frontal gyrus, dorsolateral
     \\
5/6 & ORBsup L/R     &Superior frontal gyrus, orbital part\\
7/8 &MFG L/R         &Middle frontal gyrus\\
9/10 &ORBmid L/R       & Middle frontal gyrus, orbital part\\
11/12 &IFGoperc L/R     &Inferior frontal gyrus, opercular part\\
13/14 &IFGtriang L/R    &Inferior frontal gyrus, triangular part \\
15/16 &ORBinf L/R       &Inferior frontal gyrus, orbital part\\
17/18 &ROL L/R          &Rolandic operculum\\
19/20 &SMA L/R         &Supplementary motor area\\
21/22 &OLF L/R         &Olfactory cortex\\
23/24 &SFGmed L/R      &Superior frontal gyrus, medial\\
25/26 &ORBsupmed L/R   &Superior frontal gyrus, medial orbital\\
27/28 &REC L/R         &Gyrus rectus\\
29/30 &INS L/R         &Insula\\
31/32 &ACG L/R         &Anterior cingulate and paracingulate gyrus\\
33/34 &DCG L/R         &Median cingulate and paracingulate gyrus\\
35/36 &PCG L/R         &Posterior cingulate gyrus\\
37/38 &HIP L/R         &Hippocampus\\
39/40 &PHG L/R         &Parahippocampal gyrus\\
41/42 &AMYG L/R        &Amygdala\\
43/44 &CAL L/R         &Calcarine fissure and surrounding cortex\\
45/46 &CUN L/R         &Cuneus\\
47/48 &LING L/R        &Lingual gyrus\\
49/50 &SOG L/R         &Superior occipital gyrus\\
51/52 &MOG L/R         &Middle occipital gyrus\\
53/54 &IOG L/R         &Inferior occipital gyrus\\
55/56 &FFG L/R        &Fusiform gyrus\\
57/58 &PoCG L/R        &Postcentral gyrus\\
59/60 &SPG L/R         &Superior parietal gyrus\\
61/62 &IPL L/R         &Inferior parietal, but supramarginal and angular gyrus\\
63/64 &SMG L/R         &Supramarginal gyrus\\
65/66 &ANG L/R         &Angular gyrus\\
67/68 &PCUN L/R        &Precuneus\\
69/70 &PCL L/R         &Paracentral lobule\\
71/72 &CAU L/R         &Caudate nucleus\\
73/74 &PUT L/R         &Lenticular nucleus, putamen\\
75/76 &PAL L/R         &Lenticular nucleus, pallidum\\
77/78 &THA L/R         &Thalamus\\
79/80 &HES L/R        &Heschl gyrus\\
81/82 &STG L/R         &Superior temporal gyrus\\
83/84 &TPOsup L/R      &Temporal pole: superior temporal gyrus\\
85/86 &MTG L/R         &Middle temporal gyrus\\
87/88 &TPOmid L/R      &Temporal pole: middle temporal gyrus\\
89/90 &ITG L/R         &Inferior temporal gyrus\\
\bottomrule
\end{tabular}%
}
\caption{The AAL-90 template encompasses cortical and subcortical regions along with their corresponding abbreviations and full names. Within this template, the abbreviations ``R'' and ``L'' denote the right and left brain, respectively. It is important to note that all regions are symmetrically distributed across both hemispheres of the brain.
}
\label{tab:aal90}
\end{table}

\subsubsection{fMRI Preprocessing }
\label{sec:fmriPreprocessing}
fMRI detects variations in hemoglobin levels induced by neuronal activity, altering the local magnetic field and facilitating the identification of neuronal states as either active or at rest through MRI imaging \cite{glover2011overview}.
In this study, we utilize resting-state fMRI data that has undergone meticulous preprocessing using SPM12 software \cite{chen2016resting}.
Initially, the first 10 time points are eliminated, and spatial alignment is conducted to rectify differences in data acquisition times. Subsequently, spatial normalization is executed to align the data with the Montreal neurological institute standard human brain template \cite{TZOURIOMAZOYER2002273}, and head motion correction is applied based on the initial time point. Following this, Gaussian smoothing is applied to the normalized images, and they are filtered using a low-frequency full-band filter. Ultimately, regression analysis is performed on signals from the entire brain, white matter, cerebrospinal fluid, and Friston-24 head movement parameters.

The functional connectivity network of the fMRI data was constructed utilizing a whole-brain functional connectivity analysis approach based on correlation analysis as proposed by~\citet{Salvador2005}. Specifically, we designate the 90 brain regions from the AAL template as nodes and compute Pearson correlation coefficients between the signals of these regions using the preprocessed mean fMRI signal. Subsequently, we employ the Fisher-Z transformation to produce a correlation graph represented by an adjacency matrix $\mathbf{A}^{\rm fMRI}\in \mathbb{R}^{90\times 90}$.

In summary, our dataset can be denoted as $\left \{ (\mathbf{A}^{\rm DTI}_i, \mathbf{A}^{\rm fMRI}_i, y_i)\right \}_{i=1}^N$, where $N=138$ denotes the sample size, and $y_i\in \{0,1\}$ signifies the label of the $i$-th sample. The dataset comprises $N_{\rm ASD}=67$ positive samples (ASD group) and $N_{\rm Ctr}=71$ negative samples (control group).

\subsection{ASD Diagnosis}
 The graph convolution operation $\text{GNN}(\mathbf{A}^{\rm DTI},\mathbf{A}^{\rm fMRI})$ is mainly performed with the conventional GNN algorithm, 
 \begin{equation}\label{eq:graphconvol}
     \mathbf{H}^l_i = f(\mathbf{H}_{i}^{l-1}, {\rm Aggre}\{\mathbf{H}_{j}^{l-1}|j\in \mathcal{N}(i)\};\mathbf{W}^l),
 \end{equation}
where for $l=1,\ldots, L$, $\mathbf{H}^l \in \mathbb{R}^{n\times d_l}$ is the hidden feature matrix of the $l$-th graph convolutional layer,  $\mathbf{H}^l_i$ is its $i$-th row, $\mathcal{N}(i)$ denotes the neighborhood set of node $i$, $f(\cdot,\cdot;\mathbf{W}^l)$ is a linear mapping in the $l$-th layer with parameter $\mathbf{W}^l$, and the Aggregation function, ${\rm Aggre}\{\mathbf{H}_{j}^{l-1}|j\in \mathcal{N}(i)\}$, can be any standard graph convolutional operation. We link the graph embedding $\mathbf{H}^L$ of the last graph convolution layer to a binary graph classification task, and the supervised loss function is \begin{equation*}\label{eq:trainLoss}
    \mathcal{L}_{\text{sup}} = - \sum_{i=1}^{N_{\rm train}} \left( y_i \log \sigma(\mathbf{H}^L_i)+ (1-y_i) \log(1- \sigma(\mathbf{H}^L_i)) \right) ,
\end{equation*}
where $N_{\rm train} $ is the size of the training set and $\sigma(\cdot)$ is a link function.

\subsubsection{Optimize Intra-class and Inter-class Distances}
To improve the accuracy of the graph classification, we regularize the original classification task with a term by maximizing the inter-class distance as well as minimizing the intra-class distance, as shown in Figure \ref{fig:modelFramework} (d).

Suppose the batch size is $B$, and it contains $B_1$ positive (denoted by $\mathcal{C}_1$) and $B_2$ negative (denoted by $\mathcal{C}_2$) samples, respectively. 
The within-class distance $D_k$ is defined as
\begin{equation}\label{eq:intra}
    D_k = \frac{1}{B_k(B_k-1)} \sum_{\mathcal{G}_i,\mathcal{G}_j \in \mathcal{C}_k} d^2(\mathcal{G}_i, \mathcal{G}_j), k \in \{1,2\}
\end{equation}
and the corresponding between-class distance is
\begin{equation}\label{eq:inter}
    D = \frac{1}{B_1\cdot B_2} \sum_{\mathcal{G}_i \in \mathcal{C}_1, \mathcal{G}_j \in \mathcal{C}_2} d^2(\mathcal{G}_i, \mathcal{G}_j)
\end{equation}
where $d(\mathcal{G}_i,\mathcal{G}_j)$ is a distance metric between two graphs.
We choose $d(x_i,x_j)$ to be the Wasserstein distance as defined in Section \ref{sec:wassersteindis}.

The total graph distance is 
\begin{equation}\label{eq:graphRegularization}
    \mathcal{R}_{g} = \frac{D_1 + D_2}{D},
\end{equation}
and the overall loss function of the proposed model is
\begin{equation}\label{eq:overallLoss}
    \mathcal{L} = \mathcal{L}_{\text{sup}} + \alpha \mathcal{R}_{g},
\end{equation}
where $\alpha$ is a 
tuning parameter for the graph distance regularization.

\subsubsection{Wasserstein Graph Distance}
\label{sec:wassersteindis}
Wasserstein distance serves as a graph distance computation technique that accounts for the overall structure of a graph \cite{NIPS2019_9539}. This method quantifies the distance between two graphs possessing an identical number of nodes but with unknown ordering, effectively capturing the significance of distinct edges within a complex graph structure. The Wasserstein distance evaluates the dissimilarity between two graphs $\mathcal{G}_1$ and $\mathcal{G}_2$ by leveraging their smooth graphical signals $f^{\mathcal{G}_1}$ and $f^{\mathcal{G}_2}$, formulated as
\begin{equation}
    W_2(f^{\mathcal{G}_1},f^{\mathcal{G}_2}) = \left ({\rm Tr}(L^{\dagger}_1+L^{\dagger}_2)-2{\rm Tr}(\sqrt{L^{{\dagger}/2}_1 L^{\dagger}_2 L^{{\dagger}/2}_1})\right)^{\frac{1}{2}}
\end{equation}
where $\dagger$ denotes a pseudoinverse operator, the graph signals $f^{\mathcal{G}_1}$ and $f^{\mathcal{G}_2}$ are normally distributed with a mean of zero and a covariance of the graph Laplacian matrices $L_i^{\dagger}, i\in \{1,2\}$, i.e.,
$f^{\mathcal{G}_1}\sim \mathcal{N} (0,L^{\dagger}_1),
f^{\mathcal{G}_2}\sim \mathcal{N} (0,L^{\dagger}_2).
$

Wasserstein distance emerges as a potent tool for capturing structural distinctions between two graphs, displaying heightened sensitivity to alterations in a graph's overall configuration. Notably, this distance metric gauges the dissimilarity in the prevalence of weaker graphs, presuming familiarity with the global structures of the graphs. 
We introduce Wasserstein distance as the metric $d(\cdot,\cdot)$ in Eqs.~(\ref{eq:intra}) and (\ref{eq:inter}) to evaluate the dissimilarity in spatial structures between positive and negative samples. This utilization enables us to enhance classification performance by effectively discerning spatial disparities between two classes.

\subsection{Network Node Centrality}
\label{sec:graphNodeCentrality}
\subsubsection{Degree Centrality}
Degree centrality is a measure of the importance of a node in a network based on the number of direct connections (or edges) it has to other nodes. It is a straightforward and intuitive measure, commonly used in the analysis of social networks, biological networks, and many other types of graphs. 
In an undirected network, each edge is bidirectional. The degree centrality \( C_D(v) \) of a node \( v \) is simply the number of edges connected to \( v \).

Formally, for an undirected graph \( G = (V, E) \):
\[ C_D(v) = d(v) \]
where \( d(v) \sum_{u \in V} \mathbf{A}_{vu} \) is the degree of node \( v \), representing the number of edges incident to \( v \), and where \( \mathbf{A} \) is the adjacency matrix of the graph \( G \), and \( \mathbf{A}_{vu} \) is 1 if there is an edge between nodes \( v \) and \( u \), and 0 otherwise.

To compare the degree centrality of nodes in networks of different sizes, we often normalize the degree centrality via dividing by the maximum possible degree.
For an undirected network,
\[ C_D(v) = \frac{d(v)}{|V| - 1} \]
This normalization scales the degree centrality values to the range \([0, 1]\).

Degree centrality measures the number of direct connections that a node has.
In undirected networks, degree centrality is the count of edges connected to a node.
In directed networks, degree centrality is split into in-degree (incoming edges) and out-degree (outgoing edges).
Normalized degree centrality adjusts the measure to a range from 0 to 1, allowing for comparison across different network sizes.

\subsubsection{Eigenvector Centrality}

Eigenvector centrality stands as a pivotal metric in network analysis, indicating the significance of each node's position within the overall network structure. While degree centrality in a network graph quantifies the total number of direct links that a node possesses, it fails to consider the overall network pattern. \citet{bonacich1972factoring} proposed that the eigenvector corresponding to the maximum eigenvalue of the adjacency matrix offers an effective measure of network centrality. Unlike degree centrality, which treats each connection equally, eigenvector centrality assigns weights to connections based on their centrality. This metric can be viewed as a weighted sum of both direct and indirect connections of all lengths, encompassing the entire network pattern~\cite{bonacich2007some}. 

Eigenvector centrality in a graph, often used to assign importance levels to nodes, plays a vital role in network analysis. Consider a graph $G(\mathbf{V},\mathbf{E})$ comprising vertices $\mathbf{V}$ and edges $\mathbf{E}$, with $\mathbf{A}$ representing the adjacency matrix of the graph.
We know that, 
\begin{equation}\label{eq:eigenvector}
    {\lambda}x=\mathbf{A}x, \quad {\lambda}x_{i}=\sum_{j=1}^{n} a_{ij} x_{j},
\end{equation}
where $n$ is the number of vertices. We can obtain the maximum eigenvalue ${\lambda}_{max}$ corresponding to each node, which is our sought eigenvector-centrality value. To facilitate comparison, we normalize this value to determine each node's proportional importance within the network.

\subsubsection{Subgraph Centrality}
The subgraph centrality describes the extent of participation of each node in all subgraphs of the network. It is computed through the spectrum of the network's adjacency matrix and represents a method for characterizing nodes in the network based on the number of closed walks starting and ending at the node~\cite{estrada2005subgraph}. Given a node $i$, the number of closed subgraphs of $i$ is defined as the total number of distinct closed loops starting from $i$ and returning to $i$. The smaller the path length of closed subgraphs, the tighter the connections between nodes, and the higher the corresponding centrality. The subgraph centrality of node $i$, ${\rm SC}(i)$, can be represented using the eigenvalues and eigenvectors of the adjacency matrix as 
\begin{equation}\label{eq:subgraph}
   {\rm  SC}(i) = \sum_{k=0}^{\infty}\sum_{j=1}^{N} \frac{{{\lambda}_j}^{k} (u_{ij}^2)}{k!} = \sum_{j=1}^{N}(u_{ij}^2) e^{{\lambda}_j}
\end{equation}
where ${\lambda}_j$ denotes the eigenvalues of the adjacency matrix $A$, $u_{j}$ is the eigenvector corresponding to ${\lambda}_j$, and $u_{ij}$ represents the $i$th element of the eigenvector.



 
\subsection{Two-sample Hypothesis Tests}
\label{sec:hypothesisTestDefinitions}
The disparity in network centrality between the ASD group and the control group involves a two-sample hypothesis testing procedure. Therefore, we provide a brief overview of the two hypothesis testing methods utilized in this research.

\subsubsection{Mann-Whitney U Test}
The 
Mann-Whitney U test, is a non-parametric statistical test used to assess whether there is a significant difference between the distributions of two independent groups. It serves as an alternative to the t-test when the normality assumption is not met, evaluating if one group tends to have higher or lower values than the other without assuming a specific distribution for the data \cite{mann1947test}.

\subsubsection{Maximum Mean Discrepancy (MMD) Test}
We can also utilize the MMD~\cite{bioinformatics2006,muandet2017kernel} to measure the difference between positive and negative sample distributions.
Let $p$ and $q$ denote the distributions of the eigenvector-centrality values of positive and negative samples for each node, respectively. Their squared MMD can be expressed as
\begin{equation}\label{eq:mmd_caculation}  \text{MMD}_{\mathcal{F}}^{2}[p, q] =
\sup_{{\|f\|}_\mathcal{F} \leq 1;f\in \mathcal{F}}\left\| \mathbb{E}_{p}[f(S_i^{+})]-\mathbb{E}_{q}[f(S_i^{-})]\right\|_{\mathcal{F}}^{2},
\end{equation}
where $\mathcal{F}$ is a functional space equipped with norm $\|\cdot\|_{\mathcal{F}}$.

In practical applications, the space $\mathcal{F}$ is frequently adopted as a reproducing kernel Hilbert space due to the kernel trick \cite{berlinet2004}. We use kernel mean embedding 
\begin{equation*}
\mu_{p}(\cdot)=\mathbb{E}_{p} [\mathcal K(S^+,\cdot)], \quad \mu_{q}(\cdot)=\mathbb{E}_{q} [\mathcal K(S^-,\cdot)],
\end{equation*}
where $\mathcal K$ denotes the kernel function associated with reproducing kernel Hilbert space. The 
empirical estimate of MMD \eqref{eq:mmd_caculation} is given as 
\begin{eqnarray}\label{eq:mmd_empirical}
\widehat{\text{MMD}}_{\mathcal{F}}^{2}[S^+, S^-] &=& \frac{1}{N_{\rm ASD}^{2}} \sum_{i,j=1}^{N_{\rm ASD}} \mathcal K(S_i^{+},S_j^{+}) + \frac{1}{N_{\rm Ctr}^{2}} \sum_{i,j=1}^{N_{\rm Ctr}} \mathcal K(S_i^{-},S_j^{-}) 
    \nonumber\\&&- \frac{2}{N_{\rm ASD}N_{\rm Ctr}} \sum_{i=1}^{N_{\rm ASD}} \sum_{j=1}^{N_{\rm Ctr}}\mathcal K(S_i^{+},S_j^{-}).
\end{eqnarray}
The estimator $\widehat{\text{MMD}}$  consists of U-statistics with order $2$, which can be easily implemented.

\section{Conclusion}
ASD is a prevalent neurodevelopmental disorder that poses a threat to health throughout the lifespan, often continuing into adulthood. Early diagnosis of ASD is crucial for timely intervention and treatment. Currently, the common diagnostic approach relies on questionnaires assessing external symptoms displayed by children. In recent years, researchers have started utilizing machine learning methods to assist in the diagnosis of ASD based on brain imaging data. In this paper, we propose a GNN that fuses graph data from two imaging modalities, namely fMRI and DTI. The brain connectivity networks constructed using the GNN data fusion approach enhance the diagnostic performance of ASD compared to traditional machine learning methods. Additionally, we further conduct pathological analysis based on the fused brain region connectivity graph. Specifically, we screen potential ASD pathological brain regions by analyzing the network centrality of the nodes in the fused brain region graph.
We employ two distinct hypothesis testing methods to statistically analyze the difference
in the fused networks of ASD patients and healthy control groups based on three network node centralities, identifying suspected disease-related brain regions corresponding to different centralities. Our research findings indicate variations in disease-related brain regions identified by different centrality metrics and highlight the consistency of results between the two hypothesis testing methods in assessing differences in node centrality distributions. This study offers a novel network science perspective for the etiological analysis and treatment of ASD in the later stages.









\bibliography{sn-bibliography}

\end{document}